\documentclass[conference]{IEEEtran}
\IEEEoverridecommandlockouts
\usepackage{cite}
\usepackage{amsmath,amssymb,amsfonts}
\usepackage{algorithmic}
\usepackage{graphicx}
\usepackage{textcomp}
\usepackage{xcolor}
\usepackage[a4paper, total={184mm,239mm}]{geometry}
\def\BibTeX{{\rm B\kern-.05em{\sc i\kern-.025em b}\kern-.08em
    T\kern-.1667em\lower.7ex\hbox{E}\kern-.125emX}}

\usepackage{xspace}
\usepackage{color,soul}
\usepackage{booktabs}
\usepackage{tabularx}
\usepackage{fancyhdr}
\usepackage{amsbsy}
\usepackage{pifont}
\usepackage{listings}
\usepackage[colorlinks=true, allcolors=black]{hyperref}

\definecolor{dred}{rgb}{0.8, 0.0, 0.0}
\definecolor{dgreen}{rgb}{0.0, 0.5, 0.0}
\definecolor{dyel}{rgb}{1.0, 0.77, 0.05}
\newcommand{\cmark}{\ding{51}}%
\newcommand{\xmark}{\ding{55}}%

\begin{document}




\title{Using LLMs to Facilitate Formal Verification of RTL}

\author{\IEEEauthorblockN{Marcelo Orenes-Vera,
Margaret Martonosi and
David Wentzlaff
}
\IEEEauthorblockA{Department of Computer Science and Electrical Engineering,
Princeton University\\
Princeton, New Jersey, USA\\
Email: \{movera, mrm, wentzlaf\}@princeton.edu }}

\maketitle

\pagestyle{plain}

\begin{abstract}


Formal property verification (FPV) has existed for decades and has been shown to be effective at finding intricate RTL bugs.
However, formal properties, such as those written as SystemVerilog Assertions (SVA), are time-consuming and error-prone to write, even for experienced users.
Prior work has attempted to lighten this burden by raising the abstraction level so that SVA is generated from high-level specifications.
However, this does not eliminate the manual effort of reasoning and writing about the detailed hardware behavior.
Motivated by the increased need for FPV in the era of heterogeneous hardware and the advances in large language models (LLMs), we set out to explore whether LLMs can capture RTL behavior and generate correct SVA properties.
First, we design an FPV-based evaluation framework that measures the correctness and completeness of SVA.
Then, we evaluate GPT4 iteratively to craft the set of syntax and semantic rules needed to prompt it toward creating better SVA.
We extend the open-source AutoSVA framework by integrating our improved GPT4-based flow to generate safety properties, in addition to facilitating their existing flow for liveness properties.
Lastly, our use cases evaluate (1) the FPV coverage of GPT4-generated SVA on complex open-source RTL and (2) using generated SVA to prompt GPT4 to create RTL from scratch. 
%
Through these experiments, we find that GPT4 can generate correct SVA even for flawed RTL---without mirroring design errors.
Particularly, it generated SVA that exposed a bug in the RISC-V CVA6 core that eluded the prior work's evaluation.

\end{abstract}





\vspace{-1mm}
\section{Introduction and Background}\label{sec:intro}
\vspace{-1mm}

The Cambrian explosion in the diversity of hardware caused by the end of Moore's law has exacerbated the challenges associated with RTL design verification (DV)~\cite{autocc}.
Formal property verification (FPV) utilizing industry-standard SystemVerilog Assertions (SVA)~\cite{sva_spec} is becoming increasingly important to provide exhaustive DV in the face of growing hardware complexity and variety.
SVA can express temporal relations over RTL signals, which fall into two major classes: safety and liveness properties.
Safety specifies that \textit{“nothing bad will happen”}, e.g., FSM transitions, while liveness specifies that \textit{“something good will happen”}, e.g., a request should eventually get a response.
FPV tools~\cite{jg_user,symbiotic} use solver engines based on formal methods to search for counterexamples (CEXs) exhaustively.
While FPV is very effective for DV, engineers often feel discouraged from using it because of the steep learning curve and additional effort to write assertions~\cite{formal_book}.

Prior work has tried to ease the use of FPV by automating parts of the process:
AutoSVA~\cite{autosva} generates end-to-end liveness properties from an annotated RTL module interface;
ILA~\cite{ilang} generates a model of the design from a functional specification and compares it against its RTL implementation; and
RTLCheck~\cite{rtlcheck} verifies the RTL of CPU pipelines for memory consistency by synthesizing SVA from axiomatic specifications.
While these are effective tools, they either verify subsets of RTL designs or require significant effort to write structured specifications.
With the recent advances in LLMs, a question arises: Can LLMs help accelerate RTL design and verification?
And if so, how should we integrate them into modern RTL development?
In the last few months, researchers have explored using LLMs to generate temporal logic specifications and assertions from natural language~\cite{nl2spec,sva_gen}, as well as filling gaps in incomplete RTL~\cite{rtl_gen}. 
We take a more holistic approach and \textbf{explore whether LLMs can generate correct SVA} for a given design \textbf{without any specification beyond the RTL}---even when the RTL contains bugs.
Our \textbf{motivation} for that is that specifications are not always available or precise enough. Often, implementation details are not fleshed out until the RTL is written.
This is especially true in academic and open-source hardware, where the RTL is in continuous development~\cite{ariane_git}.
Moreover, generating SVA solely from RTL would enable formally verifying RTL that has been generated by LLMs~\cite{rtl_gen}.

\begin{figure}[t]
\centering
\includegraphics[width=0.95\columnwidth]{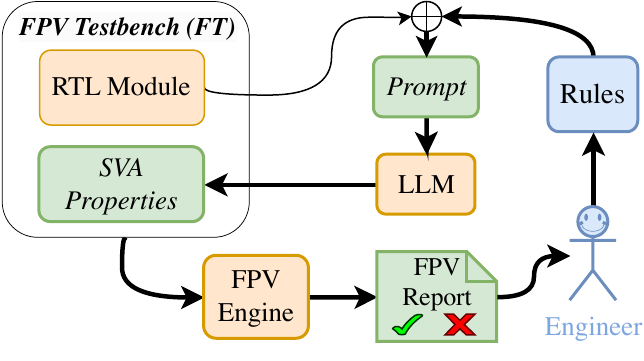}
\vspace{-2mm}
\caption{
FPV-based evaluation framework.
The FPV tool returns whether the assertions generated by the LLM are correct or not---for a given RTL.
Hinted by the errors or CEXs of the FPV report, the engineer manually writes or refines the rules that guide the LLM toward generating better SVA.
The rule set and the RTL are combined into a prompt in order to generate a new iteration of SVA properties.
The green boxes are automated steps; the blue ones are manual.
}
\label{fig:teach_loop}
\end{figure}

\textbf{Our approach:}
Starting with an empty FPV testbench (FT) for an RTL module, we aim to generate SVA that reasons about the correctness of the design without needing to manually provide details about functionality or the properties to be generated.
We utilize GPT4~\cite{openai} for this holistic task because early experiments with smaller LLMs were not promising.
However, even state-of-the-art GPT4 generates syntactically and semantically wrong SVA by default.
Thus, we first had to \textit{teach} GPT4 how to generate correct SVA by iteratively refining the prompt with rules (\autoref{fig:teach_loop}).
We then build on top of AutoSVA~\cite{autosva} to include our improved GPT4-based flow for safety properties in addition to facilitating the existing liveness flow~(\autoref{fig:autosva2}).
We make our extended framework available---anonymized for now---as AutoSVA2\cite{anonym_repo}.
Our use cases evaluate (1) the FPV coverage of AutoSVA2 on complex RTL,
and (2) using generated SVA to prompt GPT4 to create RTL from scratch, for which AutoSVA2 can output more SVA~(\autoref{fig:design_loop}).

\textbf{Our technical contributions are:}
\begin{itemize}
    \item An iterative methodology based on FPV to find the rules required to \textit{teach} an LLM how to generate syntactically and semantically correct SVA from a given RTL module.
    \item Evaluating GPT4 and crafting the rules that improve its SVA output, and extending the AutoSVA framework with this improved GPT4-based SVA generation flow.
    \item Characterizing robustness and coverage of AutoSVA2.
    \item An AutoSVA2-driven RTL generation and verification methodology; iteratively improves LLM-generated RTL by prompting human-refined generated SVA (\autoref{fig:design_loop}).
    
\end{itemize}

\textbf{Our experiments found that:}
\begin{itemize}
    \item GPT4's creativity allows it to generate correct SVA from buggy RTL, i.e., it is not compelled to generate SVA solely based on the RTL we have provided. 
    \item GPT4 is not significantly sensitive to the names of the RTL module and variables in order to generate SVA.
    \item For the same RTL modules, AutoSVA2-generated properties improved coverage of RTL behavior by up to $6\times$ over AutoSVA-generated ones.\cite{autosva_git}.
    \item Within an hour of engineering effort, our AutoSVA2 evaluation exposed a bug in the RISC-V CVA6 Ariane core\cite{ariane_git} that eluded AutoSVA's prior evaluation\cite{autosva}. 
    \item GPT4 generates better RTL when we include SVA in the prompt; its creativity allows it to generate correct RTL even if the SVA was not entirely correct.
\end{itemize}

The rest of the paper is organized as follows:
\autoref{sec:approach} shows our iterative approach to find flaws in the LLM-generated SVA and test rules to steer 
it toward generating better SVA.
\autoref{sec:autosva2} introduces AutoSVA2, our extended framework that integrates a GPT4-based flow on top of AutoSVA to create more complete FTs with less effort.
\autoref{sec:eval_ptw} and \ref{sec:design_loop} present our two uses cases.



\definecolor{codegreen}{rgb}{0,0.6,0}
\definecolor{codegray}{rgb}{0.5,0.5,0.5}
\definecolor{codepurple}{rgb}{0.58,0,0.82}
\definecolor{backcolour}{rgb}{0.97,0.97,0.96}
\lstdefinestyle{mystyle}{
    backgroundcolor=\color{backcolour},   
    commentstyle=\color{codegreen},
    keywordstyle=\color{blue},
    numberstyle=\tiny\color{codegray},
    stringstyle=\color{codepurple},
    basicstyle=\ttfamily\scriptsize,
    breakatwhitespace=false,         
    breaklines=true,                 
    captionpos=b,                    
    keepspaces=true,                 
    numbers=left,                    
    numbersep=3pt,                  
    showspaces=false,                
    showstringspaces=false,
    showtabs=false,                  
    tabsize=2,
    xleftmargin=5pt,
    framextopmargin=5pt,
    framexbottommargin=5pt,
    morekeywords={DO, NOT, USE, ONLY, MUST, SVA, \$past, ALL, TRUE, SOME, Same, Next, same, cycle, next, cycle, registers, wires, TASK, RULES, NEVER, \$countones, internal\_signal, name}
}

\definecolor{codegreen}{rgb}{0,0.6,0}
\definecolor{codegray}{rgb}{0.5,0.5,0.5}
\definecolor{codepurple}{rgb}{0.58,0,0.82}
\definecolor{backcolour}{rgb}{0.97,0.97,0.96}
\lstdefinestyle{mystyle2}{
    backgroundcolor=\color{backcolour},   
    commentstyle=\color{codegreen},
    keywordstyle=\color{blue},
    numberstyle=\tiny\color{codegray},
    stringstyle=\color{codepurple},
    basicstyle=\ttfamily\scriptsize,
    breakatwhitespace=false,         
    breaklines=true,                 
    captionpos=b,                    
    keepspaces=true,                 
    numbersep=3pt,                  
    showspaces=false,                
    showstringspaces=false,
    showtabs=false,                  
    tabsize=2,
    morekeywords={assert, property, @, cb, |, -, >, |=>, \$past, &, ==, =}
}


\section{Iteratively Teaching SVA to LLMs\label{sec:approach}}

Our early experiments with GPT4 showed us that it could generate several SVA properties solely from RTL, but they contained syntactic and semantic errors.
However, we found that we could nudge GPT4 towards generating better SVA by giving it rules to follow.
\autoref{sec:exp_setup} describes the methodology we used to iteratively construct the set of rules that are needed in the prompt for GPT4 to generate useful assertions.
\autoref{sec:exp_gpt4} describes the issues we encountered and the rules we added to the prompt to overcome them.

\subsection{Rule-refinement Methodology\label{sec:exp_setup}}

\autoref{fig:teach_loop} depicts the evaluation framework we use to iteratively refine the set of rules to be included in the prompt of an LLM.
Although all our experiments with this framework have been done on GPT4 (\autoref{table:tests}), we argue that it can be used to assess the output quality of any LLM.

This methodology requires having an FT.
We can create one quickly by executing the AutoSVA~\cite{autosva_git} script indicating the target RTL module.
(The generated FT has property and tool-binding files but no assertions.)
In theory, one could use any RTL module as input to the LLM, but note that the engineer should be able to easily determine the issues with the SVA.
Ideally, the RTL module used should be entirely correct, so that the CEXs generated by the FPV engine are due to wrong assertions.
Recall that the goal of this methodology is not to provide a complete FT for this RTL module but rather to refine the rules for the LLM to generate better SVA.
For our experiments with GPT4 (detailed in \autoref{sec:exp_gpt4}), we used the FIFO module from the AutoSVA repository~\cite{autosva_git}.

\subsection{Experiments with GPT4\label{sec:exp_gpt4}}

We apply the above methodology to GPT4 to test its generated SVA and assess whether our hand-written rules improve it.
Particularly, we used the 8K-token version offered via OpenAI's chat interface~\cite{openai}.
We use a clean-slate context for each iteration to avoid polluting the new SVA with previous issues.

\textit{\textbf{Reproducibility}}
is challenging with LLMs because of their creativity---a key attribute of our interest in LLMs for SVA generation.
Our best effort towards reproducibility is to be fully transparent with our experiments.
To do that, we made a separate commit for each iteration on our anonymized repository\cite{anonym_repo} starting from a fork of AutoSVA~\cite{autosva_git}.
This repository includes prompt, response, and FT for each iteration and a booklog with our observations for all the experiments described in this paper.

\textit{\textbf{Engineering effort:}}
It took 23 iterations and $\sim$8 hours\footnote{\label{timestampts} The engineering time can be observed from the commit timestamps\cite{anonym_repo}} to create the rules for GPT4 to output a complete and correct set of assertions for the FIFO module.
Correctness was shown by full proof, and completeness was shown from statement and toggle coverage.
We obtained assertion validity and coverage using JasperGold (JG)~\cite{jg_user}.
JG took just a few seconds to compile and test the assertions on our server.
GPT4 generated each set of SVA in under a minute.
Most of the time was spent auditing the assertions and carefully writing and refining the rules.

\textit{\textbf{LLM cost:}}
We already had a monthly subscription to this model, so these experiments did not incur extra costs.
However, note that when used via the API, queries to this model currently cost 0.03 USD per 1000 tokens, which again incentivizes using a small RTL module for this task.

\setlength{\tabcolsep}{1.5pt}
\begin{table}[t]
\vspace{-0.5mm}
\centering
\caption{Status for compilation and SVA correctness (\# of properties generated and failing) for each iteration test (T).}
\small
\begin{tabularx}{\columnwidth}{@{\hspace{5pt}}l@{\hspace{3pt}}c@{\hspace{4pt}}c@{\hspace{4pt}}c@{\hspace{6pt}}l@{\hspace{2pt}} }
\toprule
\textbf{T} & Compile & \#Prop & \#Fail & \textbf{Main issues} \\
\midrule
\textbf{1 }& {\color{dred}\pmb{\xmark}}      & 4 & $-$ & IN: Undeclared var (no module prefix) \\
\textbf{2 }& {\color{dred}\pmb{\xmark}}      & 6 & $-$ & SY: Wrong keyword for assertion \\
\textbf{3 }& {\color{dred}\pmb{\xmark}}      & 8 & $-$ & SY: Using \textit{foreach} as an assertion loop \\
\textbf{4 }& {\color{dred}\pmb{\xmark}}      & 4 & $-$ & IN: Undeclared \textit{buffer\_head\_r} \\
\textbf{5 }& {\color{dred}\pmb{\xmark}}      & 9 & $-$ & SY: Error in include and assert naming \\
\textbf{6 }& {\color{dred}\pmb{\xmark}}      & 6 & $-$ & SY: Duplicated assertion name \\
\textbf{7 }& {\color{dgreen}\pmb{\cmark}}    & 6 & \color{dred}{4} & WT: Wrong time semantics $|->$ \\
\textbf{8 }& {\color{dyel}\pmb{\xmark}}      & 6 & \color{dred}{3} & IN: Undeclared var (no module prefix) \\
\textbf{9 }& {\color{dred}\pmb{\xmark}}      & 5 & $-$ & IN: Module prefix; ignored prev. rules \\
\textbf{10} & {\color{dgreen}\pmb{\cmark}}   & 9  & \color{dred}{5} & WT: Wrong time semantics $|=>$ \\
\textbf{11} & {\color{dgreen}\pmb{\cmark}}   & 7  & \color{dred}{1} & WT: Missing \$past in postcondition \\
\textbf{12} & {\color{dgreen}\pmb{\cmark}}   & 10 & \color{dred}{7} & WT: Too much \$past usage \\
\textbf{13} & {\color{dred}\pmb{\xmark}}     & 9  & $-$ & SY: Forgot \textit{foreach} rule from T3 \\
\textbf{14} & {\color{dgreen}\pmb{\cmark}}   & 12 & \color{dred}{4} & WS: Incr. without wrap; wrong signal \\
\textbf{15} & {\color{dgreen}\pmb{\cmark}}   & 8  & \color{dred}{2} & WT: Missing \$past in postcondition \\
\textbf{16} & {\color{dgreen}\pmb{\cmark}}   & 9  & \color{dred}{4} & WT/WS: Wrong bitwise manipulation \\
\textbf{17} & {\color{dgreen}\pmb{\cmark}}   & 8  & \color{dred}{3} & WT: Wrong time semantics $|=>$ \\
\textbf{18} & {\color{dyel}\pmb{\xmark}}     & 10 & \color{dgreen}{0} & SY: Array-named assertions as\_name[i] \\
\textbf{19} & {\color{dyel}\pmb{\xmark}}     & 12 & \color{dred}{2} & SY/WS: Empty precond.;wrong bitwise \\
\textbf{20} & {\color{dred}\pmb{\xmark}}     & 10 & $-$ & SY: Wrong width in constant usage \\
\textbf{21} & {\color{dgreen}\pmb{\cmark}}   & 7  & \color{dred}{1} & WT: Missing \$past for register \\
\textbf{22} & {\color{dgreen}\pmb{\cmark}}   & 9  & \color{dred}{1} & WT: Incorrect \$past in precondition \\
\textbf{23} & {\color{dgreen}\pmb{\cmark}}   & 8 & \color{dgreen}{0} & Full Proof \\
\textbf{24} & {\color{dgreen}\pmb{\cmark}}   & 8 & \color{dred}{1} & Assuming wrong behavior about \textit{out\_rdy}\\
\bottomrule
\end{tabularx}
\vspace{-1mm}
\label{table:tests}
\end{table}

\autoref{table:tests} shows, for each iteration test (T), whether the FPV tool successfully compiled the property file, the number of assertions generated and failing, and the main issues found.\footnote{The booklog contains details of the issues observed at each iteration~\cite{anonym_repo}}

We group the issues into four categories: not compiling due to wrongly referencing internal signals (IN) or wrong syntax (SY); and compiling but using wrong timing (WT) or wrong semantics (WS).
For tests where JG could not successfully compile the FT (\color{dred}\pmb{\xmark}\color{black}), we did not attempt to fix it to check how many assertions failed ($-$), except on a few iterations when the compilation error was minor (\color{dyel}\pmb{\xmark}\color{black}).

\textit{\textbf{Internal Signals (IN):}}
Since our first iteration, we observed that GPT4 would not properly reference internal signals (not declared in the module interface).
It took several iterations to refine the rules for GPT4 to use hierarchical referencing by prefixing the module name before the internal signal.
These rules are shown in lines 4-5 of \autoref{lst:header}.

\textit{\textbf{Syntax (SY):}}
We found errors related to using incorrect keywords, e.g., \texttt{always@(<condition>)} instead of \texttt{assert}, and wrong module include and property names.
GPT4 also kept using \textit{foreach} wrongly to create loops of assertions.
Our rules to fix syntax issues are described in lines 1-3.

\textit{\textbf{Wrong Timing (WT):}}
One of the hardest issues to fix was the concept of time in SVA; e.g., GPT4 kept using same-cycle implications ($|->$) for reasoning about the updated value of registers (flip-flops). 
Timing becomes especially problematic when updating registers within an array because the index selector may also be a register.
In this case, the postcondition should reason about the updated value of the register array but the old value of the index.
We show an example of this from T16 in \autoref{lst:test_16}; the assertion should have used \texttt{\$past} for \texttt{buffer\_head\_r}.
Our rules to overcome that (lines 9-16) include teaching the concept of registers and combinational logic, same- and next-cycle assertions, and when to use \texttt{\$past}.

\textit{\textbf{Wrong Semantics (WS):}}
Beyond timing, we found other wrong semantics in counter increment logic, \texttt{\$countones} and bitwise operators.
\autoref{lst:test_16} shows an example of this, where the comment generated by GPT4 is correct, but the assertion is wrongly using \texttt{!\&} instead of \texttt{!|} to check that all buffer slots are invalid.
We fixed them by adding rules to teach GPT4 about the correct behavior of these operators (lines 6-8).

\begin{lstlisting}[escapeinside={(*}{*)}, language=Verilog, caption={
Semantically wrong assertions from iteration test 16.
}, label=lst:test_16, style=mystyle2]
// Check that if there's an in handshake (in_hsk), the buffer corresponding to the head of the FIFO should be updated with data on the next cycle.
as__in_hsk_data_update: assert property (fifo.in_hsk |=> fifo.buffer_data_r[fifo.buffer_head_r]==$past(in_data));

// Check that out_val should be low if all slots are invalid
as__out_val_low_if_all_buffer_invalid:  assert property 
    ((!&fifo.buffer_val_r) |-> fifo.out_val == 0'b0);
\end{lstlisting}

\vspace{-3mm}
\noindent
\begin{lstlisting}[escapeinside={(*}{*)}, caption={
Some of the rules we crafted to refine GPT4's SVA generation---written in plain (imperative) English based on our knowledge of SVA.
}, label=lst:header, style=mystyle]
DO NOT declare properties; DECLARE assertions named as__<NAME>: assert property (<EXPRESSION>).
DO NOT use [] at the end of assertion NAME. Do not add @(posedge clk) to EXPRESSION.
DO NOT use foreach loops in assertions, use generate for.
Internal signals are those NOT present in the interface. Internal signals are declared within the module.
Referencing internal signals in the property file ALWAYS requires prepending the name of the module before the signal name, e.g., name.<internal_signal>.
&bitarray means that ALL the bits are ONES.
!(&bitarray) means it's NOT TRUE that ALL the bits are ONES, i.e., SOME of the bits are ZEROS.
!(|bitarray) means that NONE of the bits are ONES, i.e., ALL the bits are ZEROS.
Signals ending in _reg are registers: the assigned value changes in the next cycle.
Signals NOT ending in _reg are wires: the assigned value changes in the same cycle.
USE a same-cycle assertion (|->) to reason about behavior occurring in the same cycle.
USE a next-cycle assertion (|=>) to reason about behavior occurring in the next cycle, for example, the updated value of a _reg.
DO NOT USE $past() in preconditions, ONLY in postconditions
DO NOT USE $past() on postcondition of same-cycle assertion
On the postcondition of next-cycle assertions (|=>), USE $past() to refer to the value of wires or a _reg on the cycle of the precondition.
On the postcondition of next-cycle assertions (|=>), DO NOT USE $past() to refer to the updated value of _reg.
\end{lstlisting}

\vspace{-2mm}
\subsection{Assertion Quality and Determinism\label{sec:exp_discussion}}

It took 7 iterations to get the first correct syntax and 18 iterations for all assertions to pass, but we only first achieved correct syntax and full proof at T23.

\textit{\textbf{Creativity:}}
We made an extra iteration to check for output determinism given the same input.
Of the eight assertions of T24, only two were the same as T23, and the rest were slight variations to test similar behavior.
\autoref{lst:test_23_24} shows two of the variations that check data integrity.

\begin{lstlisting}[escapeinside={(*}{*)}, language=Verilog, caption={
Assertions from T23 and 24 to check data integrity. The first one has a larger scope, since it uses more primary signals while the second one uses an intermidiate signal from the design (add\_buffer).
}, label=lst:test_23_24, style=mystyle2]
// T23: If FIFO is not empty and output handshake is true, the data at 'buffer_tail_reg' index should be sent out in the same cycle
for (genvar i=0; i<INFLIGHT; i=i+1) begin: check_data_on_wr
   as__data_correctly_written: assert property (fifo.in_hsk && !(|fifo.buffer_val_reg) && (fifo.buffer_head_reg == i) |=> fifo.buffer_data_reg[i] == $past(fifo.in_data));
end

// T24: When add_buffer flag is set, the next value of buffer_data_reg at that index should be in_data.
for (genvar i=0; i<INFLIGHT; i=i+1) begin: check_buffer_data
  as__buffer_data_set: assert property (fifo.add_buffer[i] |=> fifo.buffer_data_reg[i] == $past(fifo.in_data));
end
\end{lstlisting}

\textit{\textbf{Asserting unknown behaviors:}}
T24 had a failing assertion that checked that if the FIFO consumer is not ready to read (\texttt{out\_rdy} low), it must be because the FIFO is empty.
Since \texttt{out\_rdy} is an input to the FIFO, the FPV can set it to any value at any given time, and thus the assertion fails.
However, we found it interesting that GPT4 tried to assert the behavior of a signal that was not driven inside the RTL.

\textit{\textbf{RTL coverage:}}
With the refinement of rules, GPT4 was encouraged to use more features, and with that, it generated more assertions.
We also attribute that increase in assertion count to the better utilization of the context size.
After T9, we instructed GPT4 to output only assertions and comments, but not module interfaces or further explanations, to save tokens in favor of producing better SVA.
(\autoref{sec:autosva2} elaborates more on how to use the context efficiently.)
Via JG's coverage tool, we found that the SVA from T23 and T24 achieved full coverage of the FIFO module.
Although a FIFO is not a complex module, note that the SVA was generated from scratch solely based on the RTL code and the generic set of rules from \autoref{lst:header}.
\autoref{sec:eval_ptw} evaluates more complex RTL modules and shows how generating multiple batches of SVA increases coverage.


\textit{\textbf{Robustness:}}
We made two more tests to check the robustness of GPT4 with respect to signal names.
For T25, we replaced \texttt{fifo} with \texttt{modul} throughout the entire RTL module, including the module name, to discern whether prior knowledge about FIFO behavior enhanced GPT4's output.
GPT4 generated a set of assertions with similar quality as T24.
For T26, we replaced \texttt{modul} with \texttt{multiplier} to investigate whether prior knowledge about multipliers would worsen GPT4's output for the same RTL behavior.
Again, we found no significant impact.
We also observe no sign of GPT4 internally learning from prompting the rules repeatedly since removing the rule set results in a similar SVA as in T1.

\textit{\textbf{Rule set completeness:}}
The fact that our set of rules was good enough for our FIFO module does not mean it is sufficient for every module.
Several of the SVA features not encountered during our rule-refinement experiments could still cause trouble for GPT4.
For example, \texttt{\$countones} only appeared at T25---wrongly used---so we needed to add an extra rule.
This is to say, as more RTL modules are tested with GPT4, rules may need to be appended.
Moreover, particular strategies could be used to nudge GPT4 to generate assertions in a particular way, e.g., to write assertions about FSMs transitions (tested in \autoref{sec:expose_bug}).


\section{AutoSVA2: Extending AutoSVA with GPT4\label{sec:autosva2}}

The above experiments show that SVA generated by GPT4 is useful, not because it's perfect (which is not), but because GPT4 generates SVA that checks behaviors beyond what is written in the RTL---often spanning multiple assignment steps.

We decided to integrate our new SVA generation flow into AutoSVA because:
(1)~AutoSVA already generates the FT scaffolding we need to run SVA properties on FPV tools like JasperGold (JG)~\cite{jg_user} and YosysHQ's SBY~\cite{symbiotic}.
(2)~properties generated with AutoSVA and GPT4 can be complementary (\autoref{sec:eval_coverage});
(3)~AutoSVA is open-source and continues being extended for new features, e.g., hardware security~\cite{autocc}.

\begin{figure}[t]
\vspace{-4mm}
\centering
\includegraphics[width=0.8\columnwidth]{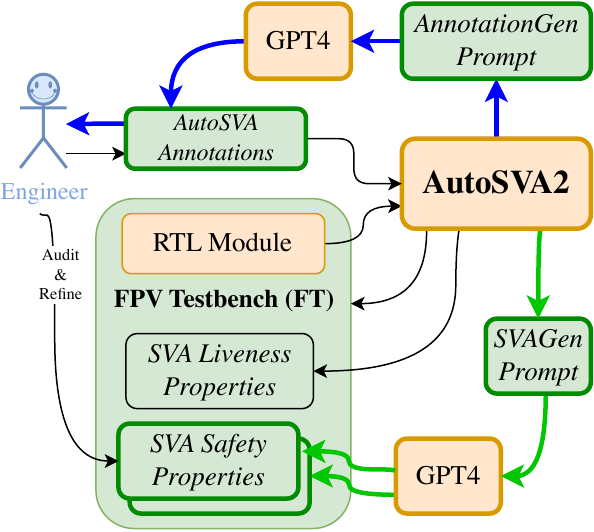}
\vspace{-1mm}
\caption{
Overview of AutoSVA2.
Our additions to the original AutoSVA flow are shown with thick boxes and arrows;
the original flow is shown with thin boxes and arrows.
The \textcolor{dgreen}{green boxes} indicate automatically generated artifacts.
The \textcolor{dgreen}{green arrows} indicate the SVA generation flow and the \textcolor{blue}{blue arrows} the annotation generation flow.
The engineer in the loop revises both the annotations and the generated SVA.
}
\label{fig:autosva2}
\vspace{-4mm}
\end{figure}

\autoref{fig:autosva2} depicts how AutoSVA2---by combining our new flow (green arrows) with the existing one from AutoSVA---creates a more complete FT
We also added an extra flow (blue arrows) to lighten the effort of adding annotations to the RTL module interface.
The engineer audits the generated annotations and SVA and corrects them if necessary.

\textit{\textbf{Interface-annotation flow:}}
We crafted another set of rules\footnote{\label{see_repo}Included in our anonymized GitHub repository\cite{anonym_repo}} to teach GPT4 how to generate the annotations about module transactions that the AutoSVA paper~\cite{autosva} introduced in order to generate end-to-end liveness properties.
As depicted in \autoref{fig:autosva2}, AutoSVA2 takes the RTL module as input and appends these rules to compose a prompt for GPT4.
We found this flow to capture transactions correctly on RTL components with clear interfaces like the FIFO, with valid syntax and semantics.
However, for more complex modules with several interfaces like those evaluated in \autoref{sec:eval_ptw}, it grouped together interfaces and signals that are not part of the same transaction, e.g., the memory request interface and the response into the TLB.
We argue that even when the annotations are incorrect, it is still easier for an engineer to correct them than starting from scratch.

\textit{\textbf{Completeness of the FT:}}
The engineer may want to generate multiple batches of SVA, as we found that to increase completeness (\autoref{sec:eval_coverage}) and having redundant assertions is not problematic for FPV tools---they can reuse the same state space exploration to prove multiple properties.
On the contrary, the annotation flow is only triggered once when the FT is first created.
The coverage metrics reported by FPV tools like JG can guide the engineer on which interface signals may be missing annotations and which other internal signals miss assertions.
Future work could extract from the FPV report the RTL lines that are not being covered and use that to generate a more targeted prompt---until the SVA covers the entire design.

\textit{\textbf{Completeness vs Correctness:}}
Having SVA with full RTL coverage does not imply that the assertions or the RTL are correct.
(a) assertions producing CEXs still contribute to coverage, and
(b) assertions may have the same bug as the RTL.
Our goal with AutoSVA2 is that the generated assertions have as much coverage as possible so that the engineer does not need to come up with new properties but rather focus on auditing the existing one to ensure it matches the hardware specification.
Note that the engineer must audit all assertions, not only those producing CEXs, as they may be proving the wrong thing.
Throughout our experiments, \textbf{we found GPT4's creativity valuable for SVA generation:} the fact that it generates similar properties with slight variations makes it prone to create buggy SVA for correct RTL but also to create correct SVA for buggy RTL.

\textit{\textbf{Filtering comments from RTL design:}}
It may seem counterintuitive at first, but we found that removing comments from the RTL design improves the quality of the generated SVA for larger RTL modules (e.g., those tested \autoref{sec:eval_ptw}).
The GPT4 model we use has a context of 8K tokens for input and output combined; when the prompt gets close to that limit, the model start forgetting part of the input in order to generate the output.

\textit{\textbf{Larger context sizes vs Modularity:}}
If the module contains too many tokens even when comments are removed ($>$500 lines), we see two options to move forward:
(1)~use a model with a larger context size, e.g., GPT4-32K, or
(2)~break down the RTL into smaller modules.
We advocate for the latter, not only because the 32K version currently costs twice as much as the 8K one per token via the API but because it is good coding practice to create submodules for self-contained functionality.


\section{Use case 1: Testing AutoSVA2 on Complex RTL\label{sec:eval_ptw}}

We applied our new SVA flow with the page-table walker (PTW) and translation look-aside buffer (TLB) of the 64-bit RISC-V CVA6 Ariane core.
We chose these modules for their complexity and because they were also evaluated in the original AutoSVA paper~\cite{autosva}, so we can compare the results.

\textit{\textbf{Goals:}}
With these experiments, we are not aiming to return fully verified RTL---auditing the CEXs would require a functional specification or more knowledge about the design.
Instead, we aim to test our GPT4-based SVA generation flow on complex RTL modules and compare the coverage of the new properties over the existing ones from AutoSVA.

\subsection{Building the PTW FT and exposing an RTL Bug}\label{sec:expose_bug}

Because CVA6 is widely used in the open-hardware community to build chips, it keeps being actively developed.
We found that the PTW code evaluated by AutoSVA has changed since then.
Tracking the bug fixes in the OpenHW Group's CVA6 repo, we found that the PTW had a bug that was fixed recently.\footnote{\label{ptw_pr}https://github.com/openhwgroup/cva6/pull/1184}
This bug was not uncovered by the assertions generated in AutoSVA's evaluation~\cite{autosva}.
Thus, we set out to evaluate whether AutoSVA2 could find this bug.

We started by rebuilding the PTW FT from the AutoSVA repository~\cite{autosva_git}.
Within a few minutes\textsuperscript{\ref{timestampts}} we had generated the first batch\footnote{We call \textbf{batch} to the output from prompting GPT4 into generating SVA.} of 12 assertions.
We generated two more batches for a total of 36 assertions.
After spending half an hour auditing the assertions and the RTL, we found one failing assertion that, if refined properly, could uncover the bug.

We kept generating more assertions to find whether GPT4 could generate the correct assertions that would fail because of the RTL bug; it did after six batches---totaling 80 assertions.\textsuperscript{\ref{see_repo}}

\begin{lstlisting}[escapeinside={(*}{*)}, language=Verilog, caption={
Assertion uncovering the PTW bug.
}, label=lst:ptw_bug, style=mystyle2]
// When in WAIT_RVALID, should remain as it was in the previous cycle if not ptw.data_rvalid_q
asgpt__wait_rvalid_tag_valid_stable: assert property (
    (ptw.state_q == WAIT_RVALID) && !ptw.data_rvalid_q |=> ptw.state_q == WAIT_RVALID);
\end{lstlisting}

\autoref{lst:ptw_bug} shows the failing assertion, which actually checks the correct transition for the PTW's FSM according to the bugfix commit.\textsuperscript{\ref{ptw_pr}}
Once we then applied the fix,\textsuperscript{\ref{ptw_pr}} the assertion proved in a few seconds of FPV tool runtime.

\textit{\textbf{Assertion generation strategy:}}
The literature on formal verification describes different strategies to write assertions~\cite{formal_book}.
For FSMs, that strategy can be creating assertions for each state transition.
We appended this to the prompt to instruct GPT4 to assert when FSMs change or retain states.
Other strategies could potentially be added to nudge GPT4 in a certain direction, e.g., to generate assertions for output signals based on intermediate signals and continue backward toward the inputs.

\subsection{RTL Coverage of Automatically Generated SVA}\label{sec:eval_coverage}

We evaluated statement and toggle coverage for PTW and TLB with different sets of assertions:
the assertions from the AutoSVA evaluation~\cite{autosva_git};
one, three, and six batches of GPT4-generated assertions; 
and all assertions combined.

\textit{\textbf{Multi-batch improvements:}}
We studied coverage improvement, starting with the existing AutoSVA assertions and adding batches of AutoSVA2 outputs.
For PTW, we obtained a 1.05$\times$, 1.23$\times$, and 1.25$\times$ increase in statement coverage with one, three, and six batches, respectively, over AutoSVA assertions alone; and 1.24$\times$, 1.57$\times$, and 1.57$\times$ increase in toggle coverage. 
For TLB, the improvements are much larger: 2$\times$, 6$\times$, and 6$\times$ increase in statement coverage; and 2.12$\times$, 3.67$\times$, and 3.74$\times$ increase in toggle coverage.
These results show that (a) AutoSVA2 improves coverage significantly over AutoSVA and (b) it is worth generating multiple batches, although we observed little to no improvements after three batches.


\textit{\textbf{Complementary assertions:}}
The assertions generated by GPT4 and AutoSVA are sometimes complementary, covering different parts of the design.
Although for the TLB we observed that having the AutoSVA assertions did not affect the final coverage, for PTW, the combination of AutoSVA and GPT4 assertions had 1.59$\times$ and 1.34$\times$ more statement and toggle coverage, respectively, over the GPT4 batches alone.
This makes sense because AutoSVA generates end-to-end properties for interface transactions, while GPT4 mostly generates assertions for internal behavior.
It is hard for GPT4 to generate end-to-end properties because they are not directly observable from the RTL.
Even for cases like the TLB where the coverage is overlapped, it is often easier for FPV tools to prove end-to-end assertions out of smaller assertions~\cite{formal_book}.


\section{Use case 2: AutoSVA2 guiding RTL generation\label{sec:design_loop}}

\begin{figure}[t]
\vspace{-3mm}
\centering
\includegraphics[width=0.94\columnwidth]{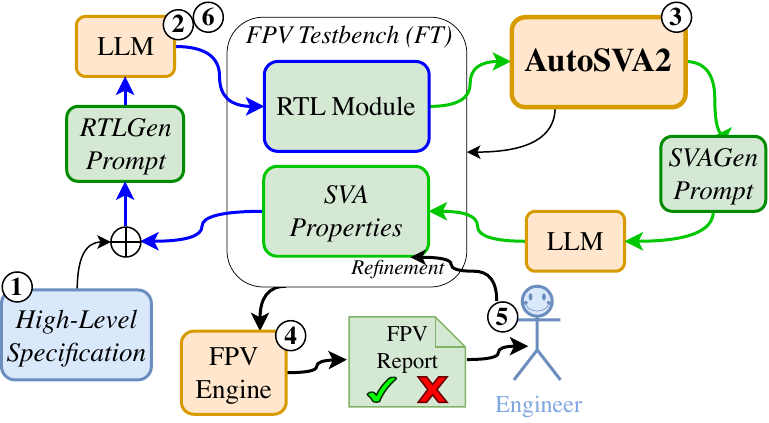}
\vspace{-1mm}
\caption{
Methodology for iteratively building RTL and FT.
The \textcolor{blue}{blue arrows} show the RTL generation flow;
\textcolor{dgreen}{green arrows}, the SVA flow;
black arrows, the assertion debugging flow with the human in the loop.
The \textcolor{dgreen}{green boxes} indicate they are automatically generated, and the \textcolor{blue}{blue box}, manually written.
}
\label{fig:design_loop}
\vspace{-3mm}
\end{figure}

LLMs have been previously used to generate RTL in partially incomplete modules~\cite{rtl_gen}.
For the holistic task of generating RTL from scratch, we set out to evaluate whether prompting GPT4 with SVA
can help it generate better RTL.

\autoref{fig:design_loop} depicts the flow we devise to generate RTL from SVA (blue arrows) in addition to the AutoSVA2 flow for generating SVA from RTL (green arrows):
{\large \ding{172}}~Start with a high-level specification in English;
{\large \ding{173}}~The LLM generates a first version of the RTL based on the specification, the module interface, and an order to generate synthesizable Verilog;
{\large \ding{174}}~AutoSVA2 generates an FT based on the RTL;
{\large \ding{175}}~JasperGold evaluates the FT;
{\large \ding{176}}~The engineer audits and fixes the SVA;
{\large \ding{177}}~The LLM generates a new version of the RTL after appending the SVA to the previous prompt.
Steps {\large \ding{174}} to {\large \ding{177}} are then repeated until \textit{\textbf{convergence}}: either (a) full proof and coverage of the FT or (b) a plateau in the improvements of the RTL and SVA.

\textit{\textbf{Our experiment:}}
We used this flow to generate a FIFO queue starting with a specification of $\sim$50 words\textsuperscript{\ref{see_repo}}).
We achieved convergence by full proof after two RTL iterations.
We discuss here our observations from this experiment.

\textit{\textbf{SVA from the first RTL:}}
From the FPV report and the SVA, we observed that 5 out of 11 assertions failed due to SVA issues; we made minor fixes in three of them (similar to T22 from \autoref{table:tests}), a partial rewrite in another one, and directly removed one that was not salvageable.
Interestingly, we found a valid assertion that failed due to a wrong RTL implementation
and an assertion that failed due to issues in both RTL and SVA regarding the empty/full flags of the FIFO.\textsuperscript{\ref{see_repo}}
We fixed that assertion but not the RTL since (as shown in \autoref{fig:design_loop}) we do not prompt the old RTL to GPT4 to generate the next RTL version.

\textit{\textbf{RTL from the refined SVA:}}
We found the RTL generated from the refined SVA to be much better than the first version; not only did it not have the bug, but the RTL was also more readable.
The assertions for the full/empty flags were still failing; we found via the CEXs that the write pointer was missing the bit selection (to ignore the carry bit) while comparing it with the read pointer.
After fixing this on the RTL, the assertion kept failing.
We observed that we had wrongly specified the empty/full flags on the manually-revised assertion from the previous step.
Once we fixed that all assertions proved.

\textit{\textbf{Takeaways:}}
From this use case, we conclude that
(a) this iterative methodology is an efficient and effective way to bring up RTL and FTs from scratch, by having the verification engineer in the loop reviewing and fixing the SVA;
(b) that errors in the RTL do not preclude GPT4 from generating correct SVA;
(c) that even if the engineer erroneously modifies the SVA, GPT4 can still generate correct RTL.





\vspace{-1mm}

\section{Discussion and Conclusion\label{sec:conclusion}}


While FPV techniques have been around for decades and are acknowledged as the most exhaustive method for DV, they are 
also exhausting for engineers to apply.
Prior work lightens this burden by raising the level of abstraction and generating SVA from high-level specifications~\cite{ilang, autosva}.
However, this does not eliminate the effort of writing specifications because, in the end, someone must reason about the detailed behavior of the hardware.
In this paper, we evaluated whether LLMs can be the ones to reason about the hardware behavior and generate, in our case, a low-level specification in SVA.


We found that GPT4---with careful guidance---can do that; it does not merely translate Verilog into SVA but rather seems to capture some of the design intent.
Integrated into the AutoSVA framework, GPT4 enables the automatic generation of FTs for RTL modules, whose completeness and correctness depend on the complexity of the design.
For small hardware components like the queue we evaluated, it requires very little human intervention to get a complete FT.

We argue that AutoSVA2 has the potential to expand the adoption of FPV---much needed in this era of heterogeneous hardware.
Moreover, producing FTs exclusively from RTL could pave the way for safer LLM-assisted RTL design approaches~\cite{rtl_gen}.
We also believe that with a curated dataset of SVA properties and their RTL, there is potential for fine-tuning LLMs to be more accurate or cost-effective (with smaller models).
Perhaps AutoSVA2 can be a starting point for generating such a dataset from open-source RTL;
\textit{\textbf{AutoSVA2 can assist you}} in generating FTs for existing RTL modules and developing new ones from scratch.
To conclude, we want to encourage the community to keep advancing and streamlining RTL design and verification methodologies that leverage the power of FPV.
Our \textit{\textbf{open-source artifacts}}~\cite{anonym_repo} include (in addition to our experiments' outputs): our rule set to guide GPT4 at generating SVA (\texttt{SVA\_GEN.v}) and AutoSVA annotations (\texttt{AUTOSVA\_GEN.v}); our template to prompt GPT4 to generate RTL (\texttt{RTL\_GEN.v}); and the script that puts everything together (\texttt{autosva2.py}).
\vspace{-1.5mm}

\bibliographystyle{IEEEtranS}
\bibliography{refs}

\end{document}